\documentclass[12pt,a4paper]{article}
\usepackage{mathtext}        
\usepackage[T2A]{fontenc}
\usepackage{amssymb,amsfonts}
\usepackage{epsf}
\usepackage{rotating}
\parskip=\baselineskip

\makeatletter

\begin{document}

\begin{center}
{\large\bf Broad-band gravitational-wave pulses\\ 
from binary neutron stars in eccentric orbits} 
\end{center}

A.V.Gusev\protect{$^{1}$}, 
V.B.Ignatiev\protect{$^{2}$}, 
A.G.Kuranov\protect{$^{1}$},
K.A.Postnov\protect{$^{1,3}$},
M.E.Prokhorov\protect{$^{1}$}

\protect{${}^1$} Sternberg Astronomical Institute, Moscow, 119899 Russia\\
\protect{${}^2$} Physical Department, Lomonossov Moscow State
University, Moscow, Russia\\
\protect{${}^3$} 
Max-Planck Institut f\"ur Astrophysik, Garching, Germany

{\it Abstract}. 
Maximum gravitational wave emission from binary stars 
in eccentric orbits occurs near the periastron passage. 
We show that for a stationary distribution of binary neutron stars
in the Galaxy, several high-eccentricity systems with
orbital periods in the range from tens of minutes to several days
should exist that emit broad gravitational-wave pulses
in the frequency range 1-100 mHz. The space  
interferometer LISA could register the pulsed signal
from these system at a signal-to-noise ratio level
$S/N>5\sqrt{5}$ in the frequency range       
$\sim 10^{-3}-10^{-1}$ Hz during one-year observational
time. Some detection algorithms for such a signal are
discussed.

\section{Introduction}

The operation of the first broad-band laser
interferometric antennas (LIGO, VIRGO, GEO-600) is coming up 
The sensitivity of theses detectors in the frequency range 10-1000 Hz 
will be sufficient to observe astrophysical sources of
gravitational waves (Braginsky 2000 and refs. therein). 
It is widely accepted that compact binary stars 
consisting of neutrons stars (NS) and black holes (BH) 
are among the primary sources for these detectors (see
Grishchuk et al. 2001 for a recent review). 
In the high-frequency band, which the first
interferometers are designed for
\footnote{We remind that the sensitivity of the 
ground-based detectors is limited at low frequency by 
seismic noises}, the spiral-in phase of the 
binary coalescences and their merging phase
are to be observed, which can in principle be
associated with spectacular astrophysical phenomena 
with an electromagnetic energy release of order of
the NS binding energy (for example, with cosmic gamma-ray
bursts, as was suggested by Blinnikov et al. 1984).   

In the case of two material points of equal mass in a circular orbit
with period $P$, the coalescence time on the quadrupole
approximation is approximately  
$$
t_0\sim 10^5 (P/[1\hbox{s}])^{8/3}\,[\hbox{s}]
$$
i.e. about 20 minutes for the initial 
$P=0.2$ s corresponding to a GW frequency of 10 Hz
at the low-frequency sensitivity boundary  of the
ground-based interferometers. The signal has a quasi-periodic
character with gradually increasing amplitude and frequency
as the stars approach each other (the so-called ``chirp'' signal). 
The waveform of this signal can be calculated
with a very high accuracy so that a powerful optimal
filtering technique
(well known in radiophysics, 
e.g.. Tikhonov 1966) can be applied to extract
the signal out from the instrumental noise
(see Grishchuk at al. 2001 for more references).
  
At lower frequencies, the searching for GWs is possible 
using space antennas. The most advance project to date 
is the LISA interferometer (Bender et al. 2000). The planned
sensitivity range for this detector lies within  
$10^{-4}-10^{-1}$ Hz. 
Main sources for LISA include massive binary BH coalescences
in remote galaxies, individual close binaries in
our Galaxy, and, possibly, a relic GW background 
originated in the very early Universe (Grishchuk et al. 2001).
Galactic close compact binaries (white dwarfs, NS, BH) 
mainly compose a stochastic GW noise at frequencies
below 1 mHz. The binary stochastic GW background
and the possibility of detecting individual binaries 
with compact objects by LISA 
was recently reconsidered by Nelemans et al. (2001).

In this letter, we study GW-signals from high-eccentricity 
binary NS, which appear as a sequence of broad-band 
pulses with a period from tens of minutes to several
hours in the frequency range $10^{-3}-10^{-1}$~Hz. These pulses 
form during the periastron passages of binary 
NS, where most GW are emitted.
For a stationary galactic binary NS coalescence rate
of $10^{-4}$ events per year, the number of binary NS
that can produce such pulses at frequencies 1-100 mHz 
at the LISA signal-to-noise level $5\sqrt{5}$ is
estimated to be around 7 for one-year observation time. 
   
\section{Stationary distribution function of binary neutron
stars in the Galaxy}

Binary NS (and BH) are natural products of evolution of
massive stars. In the case of NS, core-collapse type II or Ib/c
supernova explosions occur twice during the evolution of 
the original massive binary. It is well known (Blaauw 1961) that 
the sudden mass-loss from one component of a circular binary system
due to supernova explosion makes the orbit non-circular
with an eccentricity   
$e=\Delta M/M'$, where $\Delta M$ is the mass lost
and $M'$ is the total mass of the system after
the explosion. If an additional kick velocity 
is imparted to the newborn NS (which largely follows
from radio pulsar space velocity observations, e.g.
Lyne and Lorimer 1994, Arzoumanian et al. 2001, or 
is directly indicated by the non-coaxiality between
the NS spin and the orbital angular momentum in some 
binary pulsars, e.g. PSR B1259-63, Prokhorov and Postnov 1997, 
PSR J2051-0827, Doroshenko O. et. al 2001),
the resulting binary eccentricity become even larger on
average. 

In addition to having a sufficiently large amplitude, a GW signal
should be frequent enough to have chances to be detected
during a reasonable observational time (usually taken to be 
one year). At present, a good theoretical estimate of the 
galactic binary NS coalescence rate is $10^{-4}-10^{-5}$
per year (Lipunov et al. 1997, Yungelson
and Portegies Zwart 1998 and later papers, see 
Grishchuk et al. 2001 for a more detailed discussion).
The main theoretical uncertainty pertinent to 
such estimates is due to unknown anisotropy of
the stellar core collapse. Since only close system
that evolve rapidly enough can coalesce due to GW emission
over the galactic age $\sim 10^{10}$ years, the galactic 
birthrate of binary NS must be higher. So it is expected that 
under stationary conditions $N=10^6$ binary NS can 
simultaneously exist in the Galaxy. 

Assuming that GW emission is the only mechanism 
for the orbit decay (which is a very good approximation
in sufficiently low-dense stellar fields; in dense
stellar systems dynamical interactions should be taken into 
account) and that a stationary star formation has
really taken place in the Galaxy over past 5 billion years,
one can calculate a binary NS distribution function 
over orbital parameters (orbital large semi-axis $a$
and eccentricity $e$), which is normalized as 
$N=\int dN(a,e)da\,de$. 

This problem was solved for model initial parameters by
Buitrago et al. (1994). In our previous paper (Ignatiev et al. 2001), 
the distribution function $dN(a,e)$ was found for the initial 
orbital parameters of binary NS computed using the population
synthesis code "Scenario Machine" (Lipunov et al. 1996). In the 
parameter
region where the binary coalescence occurs over the time
interval shorter than the galactic age, the distribution function 
becomes stationary. A fragment of the stationary 
distribution calculated using the Maxwellian distribution of
the kick velocity with a mean value of 200 km/s is shown in Fig. 1. 
The normalization is made for the galactic 
binary NS coalescence rate  $10^{-4}$. Clearly, the function 
sharply increases at low orbital frequencies (large orbital
periods). The shape of the initial distribution
function has most effect exactly in this region.             
 
\section{GW-signal from binary NS at periastron passage}

A binary star with the components of mass 
$M_1$ and $M_2$ in a non-circular orbit emits gravitational
waves in a wide frequency range on harmonics to the orbital 
Keplerian frequency  
$(2\pi \nu_K)^2=G(M_1+M_2)/a^3$ (Peters and Mathews 1963).
The main energy is emitted when the stars approach the periastron 
distance $a_p=a(1-e)$. 
For large eccentricities, the duration of the GW pulse is approximately 
$\tau\sim a_p/v_p$ ($v_p$ is the velocity at the periastron)
and the maximum of radiation is emitted in the harmonic
whose number
$n\sim(\tau\nu_K)^{-1}\sim (1-e)^{-3/2}$ in accordance with the
3d Kepler's law. 

Let us calculate more precisely the time of the most 
energy emission during the periastron passage. We shall use 
the well known expression for GW emission from a binary in elliptical
orbit averaged over the binary orbit orientation
(Peters and Mathews 1963, Landau and Lifshitz 1988)
\begin{equation}
-\frac{dE}{dt}=\frac{8G^4M_1^2M_2^2(M_1+M_2)}{15a^5c^5(1-e^2)^5}
(1+e\cos\phi)^4[12(1+e\cos\phi)^2+e^2\sin^2\phi]\,.
\end{equation}
Here and below 
$G$ is the Newton gravitation constant, $c$ is the speed of light, 
$\phi$ is the polar angle counted from the periastron 
Next, we define the periastron GW pulse duration $\tau$ as the 
time of emission of 90\% of the energy released over one
orbital period  $P$. Let us change integration over time $t$ 
by integration over the angle $\phi$
$$
d\phi=\frac{(1+e\cos\phi)^2}{(1-e^2)^{3/2}}\sqrt{\frac{G(M_1+M_2)}{a^3}}
dt
$$
(we assume that GW-induced change in the orbital energy and angular momentum
is small and does not affect the Keplerian parameters of the orbit; 
this is justified for orbital frequencies considered).
The dependence of the polar angle $\phi_{90}$
that determines the corresponding part of the orbit 
is shown in Fig. 2. 
The plot $\tau/P$ versus eccentricity is 
also shown in Fig. 2. Starting from $e\approx 0.5$
this angle is about 60 degrees and slowly changes with  
eccentricity (which is intuitively clear: 
at large $e$ the focal parameter of ellipse
$p$ and the periastron distance $a_p$ are both 
proportional to $(1-e)$ and the periastron parts of
orbits with different $e$ are similar). As also follows from Fig. 2, 
already at $e\sim 0.3$ 90\% of the energy is released over 
1/3 of the orbital period. Below we shall take $e=0.3$ as
a low boundary of binary NS eccentricities which can
produce broad GW pulses.  

Thus, a binary NS in an elliptical orbit emits a GW signal
in a wide frequency range $\Delta \nu_b\sim 1/\tau$. 
The pulses are characterized by the duration $\tau$, 
the recurrent period $P$, and by some amplitude $h$.
The latter cane be estimated as a maximal dimensionless
GW amplitude $h_{max}$ at the periastron. Averaging the squares
of the field amplitudes $h_+$ and $h_\times$ at the periastron ($\phi=0$)
(Peters and Mathews 1963, Moreno-Garrido et al. 1995) over
the binary star orbit orientation with respect to the line of sight, 
we arrive at 

\begin{equation}
\label{h_max}
\begin{array}{cl}
h_{max}&=\left.\sqrt{h_+^2+h_\times^2}~ \right|_{\phi=0}=\\[3mm]
&=\sqrt{\displaystyle\frac{32}{5}}\displaystyle\frac{G^2M_1M_2}{c^4a(1-e^2)}
\displaystyle\frac{1}{r}
(1+3e+\frac{10}{3}e^2+\frac{5}{3}e^3+\frac{1}{3}e^4)^{1/2}\,, \\
\end{array}
\end{equation}
where $r$ is the distance to the system.
In the limit $e\to 1$, this maximum amplitude 
exceeds by $\sqrt{28/3}\approx 3$ times the amplitude
of GW in a circular orbit with radius equal to 
the focal parameter $p$, and by $\sim 1.5$ times
that with the periastron distance. We also note that 
the mean amplitude   
$\sqrt{h_+^2+h_\times^2}$ for the case of $\tau_{90}$
changes with increasing eccentricity at the endpoints of the corresponding
time intervals 
by no more that 2 times, and by no
more than 4.5 times for the case of $\tau_{99}$
(Fig. 2, the bottom panel). 

For numerical estimates we should know the distance to the source. 
Since during a binary NS formation the barycenter of the system 
can acquire an appreciable space velocity, the galactic subsystem
of binary NS has a larger size than that of ordinary stellar components. 
Quantitative calculations were performed by different authors
(e.g. Bulik et al. 1999), so we shall use their results. 
For example, for the kick velocity 200 km/s the mean distance to 
a binary NS from Sun is about 12.15 kpc. For numerical estimates we
shall assume all binary NS stars to lie at this average distance.

\section{Detecting broadband GW signals}

To calculate the expected detection rate 
of GW signals considered it is necessary to precise 
methods of their possible detection. First, let us
choose the frequency range in which the signals 
are to be searched for. It is known (see Grishchuk et al 2001,
Nelemans et al 2001) that the detection of GW signals in the
LISA frequency band up to $\sim 1$ mHz is hampered\footnote{But is not totally
excluded, see the  resent paper by Hellings (2001)} 
by the unavoidable presence of a stochastic noise formed
by unresolved galactic binary white dwarfs (the 
stochastic background associated with unresolved 
galactic binary NS even with account of non-circular
systems is several times lower due to much lower
amount of binary NS than white dwarfs, see Ignatiev et al. 2001).
So in what follows we shall consider only 
the frequency range $10^{-3}-10^{-1}$ Hz.

Binary NS in elliptical orbits whose GW emission at 
periastron falls within this frequency range have 
orbital periods $P$ from tens of minutes to several
days (in the case of very large eccentricities). That is, 
during one year observation time ($T\approx 3\times 10^7$ s) 
the signal from one system will consist of a periodic sequence
with $k=T/P$ broad pulses with an arbitrary (from the 
beginning of the observation) phase. The direction to
the source is also unknown. Clearly, there should be more than 
one such systems (cf. Nelemans et al. 2001). So from the viewpoint of
detection the total signal is a superposition of 
periodic pulse sequences $s_i(t)$ with arbitrary phases 
and different periods
$s(t)=s_1(t)+s_2(t)+\ldots+s_n(t)$. 
We stress that in the frequency range considered such
a signal does {\it not form} a continuous background
with the detector frequency resolution  
$\Delta \nu=3\times 10^{-8}$~Hz (Ignatiev et al. 2001)
and the planned LISA sensitivity $S_n$.

Consider now an ideal case where there is only one sequence of such pulses.
The signal-to-noise ratio when observing one broad pulse with 
the duration $\tau$ by a detector with a sensitivity $S_n$ [Hz$^{-1/2}$] is
\begin{equation}
\label{S/N_1}
(S/N)_1=\frac{h_{max}}{\sqrt{\Delta \nu_b S_n}}
\end{equation}
where $\Delta\nu_b=1/\tau$  is the pulse frequency width.
For a periodic sequence of $k$ identical pulses 
we obtain  
\begin{equation}
\label{S/N_k}
(S/N)_k=(S/N)_1\sqrt{k}=\frac{h_{max}}{\sqrt{\Delta \nu S_n}}
\sqrt{\frac{\tau}{P}}
\end{equation}
where $\Delta \nu=1/T=3\times 10^{-8}$~Hz is 
the maximum frequency resolution over the continuous 
observation time. 

As an example, consider a pulsar similar to the Hulse-Taylor's one
PSR B1913+16 with orbital parameters (Taylor and Weisberg 1989)
$a=1.95\times 10^{11}$~cm, $P=27907$~s, $e=0.617$, 
the equal component masses 
$M_1= M_2=1.4 M_\odot$, located 
$\simeq 5$~kpc away. 
Averaging over the binary orientation, 
we find $h_{max}\approx 10^{-22}$. With the eccentricity 
0.617 the pulse duration is
$\tau_{90}\approx 0.15 P=4186$~s, 
$\Delta\nu_b\approx 0.23$ mHz, the frequency 
 of the maximal (4th) harmonic is 
$\approx 0.18$ mHz, and the signal-to-noise ratio
assessed using Eqn (\ref{S/N_k}) is only 0.5
(without taking into account the binary 
white dwarf noise in this range). Clearly,
such sources are of no interest for us. 

For reliable detection of a pulse signal coming from unknown direction
one usually requires $S/N>5\sqrt{5}$ where the factor $\sqrt{5}$ 
accounts for unknown coordinates of the source (Thorne 1987). 
With such a high detection threshold the integration 
of the stationary distribution function over
the region where binary NS systems giving such a high
amplitude reside is $\sim 7$. Clearly, the lower the detection
threshold the larger the number of the system (for example, 
it is about 15 for $S/N=5$). Increasing the pulse duration 
(e.g. taking $\tau_{99}$ instead of $\tau_{90}$) could formally 
somewhat increase the number of detected systems due to 
decreasing the frequency band $\Delta\nu_b$ in Eqn (\ref{S/N_1}),
however in this case the systems should be chosen
starting from a higher eccentricity ($\sim 0.5$) in order that
the pulse duration be less than 1/3 of the orbital period.
So in fact the corresponding region of integration of the
distribution function gets narrower and the number
of detections  decreases.  

The distribution of binary NS for the stationary distribution function
under consideration over the pulse widths $\tau$ and orbital periods $P$ is
shown in Fig.  \ref{dn_dtau}. The maximum probability is to detect 
pulses about 100 s in duration from systems with orbital
period near 1000 s. The mean eccentricity of such systems, 
as follows from Fig. 2, the middle panel, is about 0.5.

Thus, actually several systems with sufficiently high amplitudes 
simultaneously contribute to the total signal. Here we
meet the typical problem of parameter estimation of a quasi-determined
signal in the background noises. For Gaussian noises and the
a priori known number of sources this problem can be solved
exactly using the well known methods of
signal detection in radiophysics. The basic principles 
of constructing optimal detectors for quasi-determined signals
with Gaussian noises are collected in the Appendix.
 
\section{Discussion}
 
1. {\it Stationarity} of binary NS distribution 
in the Galaxy over orbital parameters is one of the 
main assumptions used. This assumption is justified when the 
characteristic evolutionary time scale of the systems 
is much shorter than the time of significant 
variations of the initial source distribution function.
We assume that the star formation rate in the Galaxy can be 
considered constant at least over the last 5 billion years 
(a different assumptions is used in the paper by Nelemans et al. (2001)).
The evolutionary time scale of compact binaries in the 
frequency range considered $10^{-5}-10^{-1}$~Hz is 
much shorter than this time interval so the stationarity 
can be established. The increase of the star formation rate
in the past leads to some increase of the value of the 
distribution function, especially at low frequencies.
However, parameters of binary NS formation (especially 
the value of the kick velocity and its distribution) 
has more effect on the actual binary NS distribution. 

2. {\it Normalization} of the distribution function 
to provide the binary NS coalescence rate 
$10^{-4}$ per year is a good estimate which is
close  to the {\it upper} accepted boundary of
such evaluations (Grishchuk et al. 2001 and references therein).
Changing this normalization would lead to the corresponding change in 
the resulting numbers. 

3. We {\it do not} consider a more numerous potential class of
sources like a NS + a massive white dwarf. Systems in circular
orbits are outside our scope, and systems NS+massive WD in elliptical orbits
(studied for example by Portegies Zwart and Yungelson (1999)) 
in the frequency range of interest are not distinguishable 
from double NS systems. A detailed analysis of formation of 
binary systems comprising an old white dwarf and a young NS,
thereof representatives can be binary radio pulsars 
PSR B2303+46 and PSR J1141-6545, has recently been performed
by Tauris and Sennels (2000). According to these authors, 
the (model-dependent) formation rate of such systems in the Galaxy
can substantially exceed that of binary NS (see also 
Brown et al. 2001). We also omit compact binary NS+BH
and BH+BH systems due to their small number in comparison
with binary NS. Anyway, the possible contribution of systems
like NS+WD, NS+BH, BH+BH in non-circular orbits will
increase the number of pulses analyzed.  

\section{Conclusion}
We have analyzed the broad pulse GW signal produced by 
compact binary stars during the periastron passage. 
In the particular case of a stationary distribution of
binary NS in the Galaxy over orbital periods and
eccentricities normalized to the galactic coalescence rate
$10^{-4}$ per year, in the frequency range $10^{-3}-10^{-1}$~Hz, 
one can expect to detect with the LISA space interferometer 
$\sim 7$ individual sources 
at the signal-to-noise ratio level $5\sqrt{5}$. Allowing
for other types of compact binaries (NS with white dwarfs 
or black holes) can increase this number. 
The GW signal from each system represents a sequence of
broad pulses with equal amplitude. The most plausible
width of the pulses is around 100 s, the period (the orbital period
of the underlying binary system) is around 1000 s. From th viewpoint
of signal detection theory, some optimal detection 
algorithms of such a quasi-determined signal in the background of 
additive noise are discussed. 

The authors acknowledge L.R.Yungelson for constructive notes. 
The work is partially supported by the RFBR grants 00-02-17164,
00-02-17884a, and 01-15-88310m. KAP thanks the MPA f\"ur 
Astrophysik (Garching) for hospitality. 
  
\section{Appendix}

Let
 $$
 x(t)=S(t,\Theta_1,...,\Theta_N)+n(t)~,\quad 0 \leqslant t \leqslant T
 $$
be an additive mixture of a quasi-determined signal═$S(t)$
 $$
 S(t,\Theta_1,...,\Theta_N)=\sum\limits_{i=1}^N \Theta_i S_i(t)~,
 \quad 0 \leqslant t \leqslant N
 $$
where $\Theta=(0,1)$ is the detectability parameters, 
and a Gaussian noise $n(t)$ with the known correlation function
$K_n(\tau)$. 
 \smallskip

 {\it A. Detectability parameters are known}.
The hypothesis that the signal is present is accepted if the
following condition is met (e.g. Sosulin 1992)
  $$
  Z(\vec Y,\Theta_1,...,\Theta_N)=
 \sum\limits_{i=1}^N \Theta_i Y_i- 
 \frac{1}{2}\sum\limits_{i=1}^N\sum\limits_{k=1}^N 
 \Theta_i \Theta_k \rho_{ik}> C_{\alpha}~,
 \eqno{(A1)}
  $$
  Here
  $$Y_i=\int\limits_0^T x(t) u_i(t)dt~,\quad 
  \rho_{ik}=\int\limits_0^T S_i(t)u_k(t)dt~,
  $$
 $$
  \vec Y=(Y_1,...,Y_N)~.
  $$
 where the fiducial signal  $u_i(t)$, $i=\overline{1,N}$ 
is the solution of the type I Fredholm equation
   $$
  \int\limits_0^T K_n(t-\tau)u_k(\tau)d\tau =S_k(t)~,\quad
  0 \leqslant t \leqslant T~.
  $$
The threshold level $C_{\alpha}$ depends upon the 
false alarm probability  
  $\alpha$ (Neumann-Pierson test).

The analysis of Eqn (A1) shows that for known detectability
parameters the optimal detector represents an $N$-channel linear
system. The principal element of an individual system is a correlation 
detector that from the independent variable $Y_i$, $i=\overline{1,N}$
(in the filtration variant the correlation detector is substituted by
an optimal filter).

  {\it B. Detectability parameters $\Theta_i$ are unknown.
Bayesian algorithm of detection}.
Consider the detectability parameters  $\Theta_i$ to be independent 
discrete random values with known distribution 
  $$
  P\{\Theta_i=1\}=p_i~,\quad P\{\Theta_i=0\}=1-p_i~,\quad i=\overline{1,N}~. 
  $$
It can be shown that under the condition of parametric apriori uncertainty
the hypothesis that the quasi-determined signal  
  $S(t,\Theta_1,...,\Theta_N)$ is present can be accepted if 
  $$
   Z(\vec Y,\hat\Theta_{1B},...,\hat\Theta_{NB})>C_{\alpha}~,
  \eqno{(A2)}
  $$
   where  $\hat\Theta_{iB}$ is a Bayesian estimate 
of the unknown parameter with a quadratic payment function
 $$
\hat\Theta_{iB}=
=\underbrace{\int...\int}_{N}
\Theta_i\exp\left\{Z(\vec Y,\Theta_1,...,\Theta_N)\right\}
W_{pr}(\Theta_1,...,\Theta_N)d\Theta_1...d\Theta_N~,
 \eqno{(A3)}
$$

 $$
 W_{pr}(\Theta_1,...,\Theta_N)=\prod\limits_{i=1}^N
\left[(1-p_i)\delta(\Theta_1)+p_i\delta(\Theta_i-1)
\right]~.
 $$
is an apriori joint probability density of random variables  
$\Theta_1,...,\Theta_N$.
 \smallskip
 
 {\it C. Detectability parameters are unknown. Non-Bayesian 
detection algorithm}. For non-parametric apriori uncertainty
the detectability parameters  $\Theta_i$ are considered as
non-random. When constructing a detection device, parameters  $\Theta_i$
are substituted by maximum likelihood estimates  $\hat\Theta_i$,
which leads to the following discrimination rule:
$$
 Z(\vec Y,\hat\Theta_{1},...,\hat\Theta_{N})>C_{\alpha}~.
 \eqno{(A4)}
$$
The maximum likelihood estimates $\hat\Theta_i$ are chosen such
that 
$$
 Z(\vec Y,\hat\Theta_{1},...,\hat\Theta_{N})>
 \max\limits_{\Theta_1,...,\Theta_N} Z(\vec Y,\Theta_1,...,\Theta_N),
 \eqno{(A5)}
$$
The analysis of Eqn (A2-A5) indicates that:

1) For detecting quasi-determined signal 
 $S(t,\Theta_1,...,\Theta_N)$ the detector should contain an
additional unit, the estimator of unknown detectability parameters  $\Theta_1,...,\Theta_N$.
 
2) Detection algorithms (A2) and (A5)
prove to be {\it non-linear}:
 $\hat\Theta_i=\hat\Theta_i(\vec Y)$,~ 
 $\hat\Theta_i=\Theta_i(\vec Y)\ne \hat\Theta_{iB}$.

3) For a non-Bayesian detection algorithm, the block of 
estimates of unknown detectability parameters represents a multi-channel
linear system (see (A5)). The output signal from this 
systems is used in the matching scheme. 
 
4) The detection of the quasi-determined signal 
 $S(t,\Theta_1,...,\Theta_N)$ using algorithms (A2) and  (A4)
can be considered as a joint detection and measurement of
parameters $\Theta_i$ of the original signal.
 
5) The structure of the Bayesian estimation block is essentially
non-linear so such system are not widely used in practice. 
 
The threshold level  $C_{\alpha}$ in Eqn (A2) and  (A4)
is determined by the following conditions
$$
  \left.
  \begin{array}{l}
  P\{Z(\vec Y,\hat\Theta_{1B},...,\hat\Theta_{NB})>
  C_{\alpha}|_{\Theta_1=\Theta_2=...=\Theta_1=0}      
  \}=\alpha~,\\[3mm]
  P\{Z(\vec Y,\hat\Theta_{1},...,\hat\Theta_{N})>
  C_{\alpha}|_{\Theta_1=\Theta_2=...=\Theta_1=0}      
  \}=\alpha~,\\
  \end{array}
  \right\}
  \eqno{(A6)}
$$
where in Eqn (A6)~
 $\hat\Theta_{iB}$ and $\hat\Theta_{i}$ are pseudo-estimates
of unknown detectability parameters in the absence of the 
signal.
 
Since algorithms (A2) and (A4) are non-linear, distribution functions (A6) 
can be only empirically found from a computer simulation of
Gaussian noises with a given correlation function
(given spectral density).

\newpage
{\large Figure Captions}

\begin{figure}
\centerline{\epsfysize=1\hsize
\epsfbox{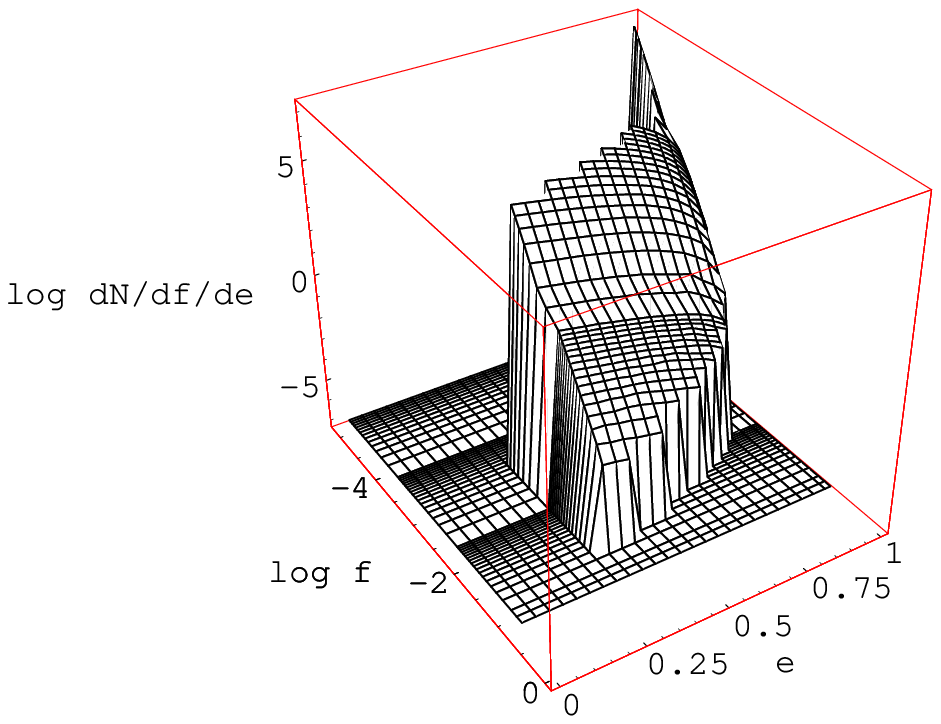}}
\caption{ }
\label{dn/df/de}
\end{figure}

Fig.~\protect\ref{dn/df/de}.
The stationary distribution function for binary neutron stars
in the Galaxy
$dN/df/de$ calculated in 
Ignatiev et al. (2001) for Maxwellian kick velocities
with the mean value 200 km/s. The normalization to the
galactic binary NS coalescence rate
$10^{-4}$ per year. Shwn is a fragment for systems with 
 $e>0.3$ which coalesce over the time interval longer than 1 year
(the upper boundary on frequencies and eccentricities) and produce 
a GW amplitude in the periastron $h_{max}$ estimated from Eqn 
(\protect\ref{h_max}). The signal-to-noise ratio is
$5\sqrt{5}$ for one sequence of pulses  
(\protect\ref{S/N_k}).

\begin{figure}
\centerline{\epsfysize=1\hsize
\epsfbox{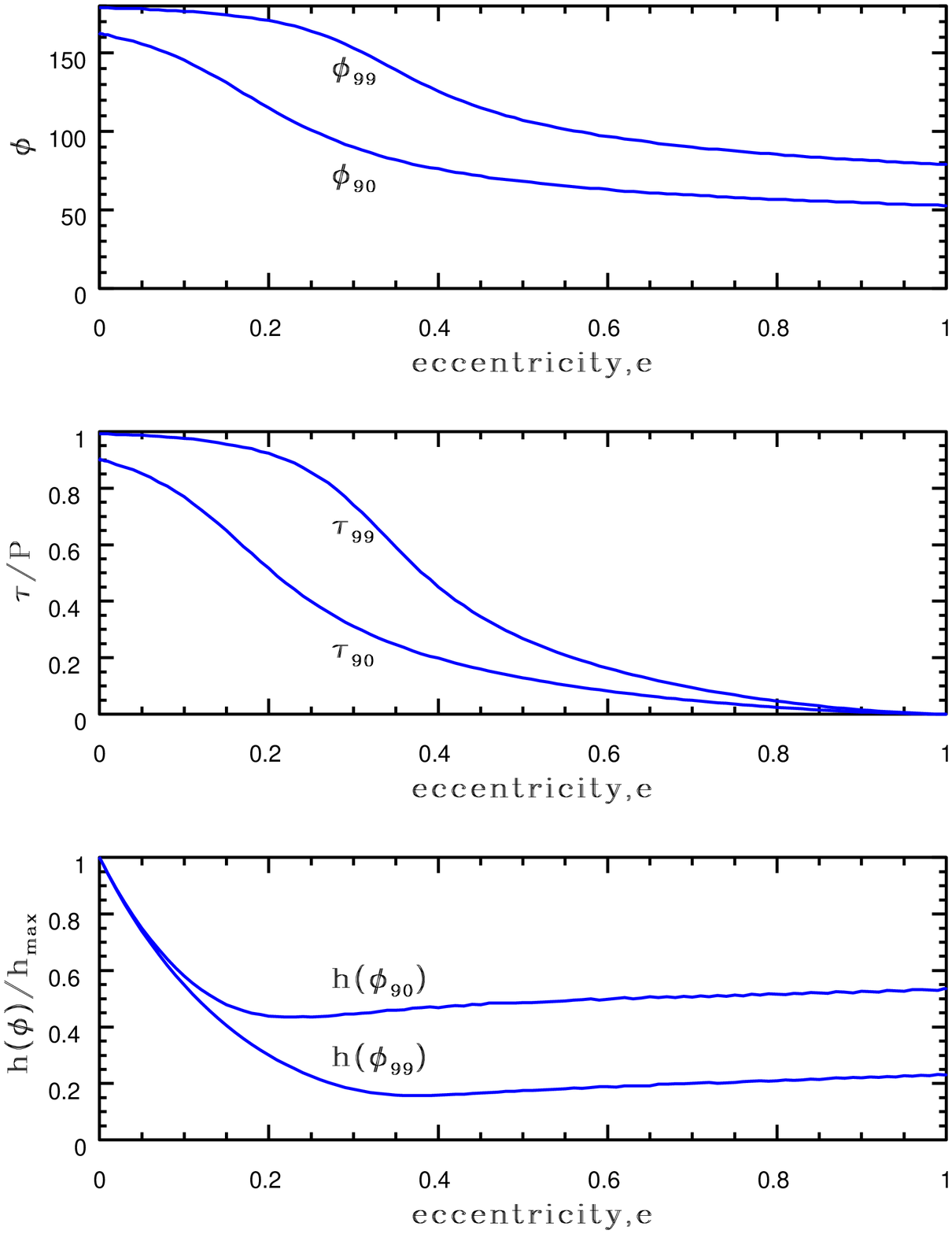}}
\caption{ }
\label{t_e}
\end{figure}

Fig.~\protect\ref{t_e}.~
Upper panel:
the polar angle $\phi$ of an elliptical orbit 
inside which $99\%$ ($|\phi|<\phi_{99}$) and 90\% 
($|\phi|<\phi_{90}$) of energy is emitted 
in the periastron passage, as a function of $e$.
Middle panel: the pulse duration 
$\tau_{99}$ and $\tau_{90}$ 
in units of the orbital period $P$. Bottom panel:
the variation of the mean amplitude 
$h(\phi)/h_{max}$ within the boundary points of the orbit
$\phi_{90}$ and $\phi_{99}$.

\begin{figure*}
\begin{turn}{-90}{
\centerline{\epsfysize=1\hsize
\epsfbox{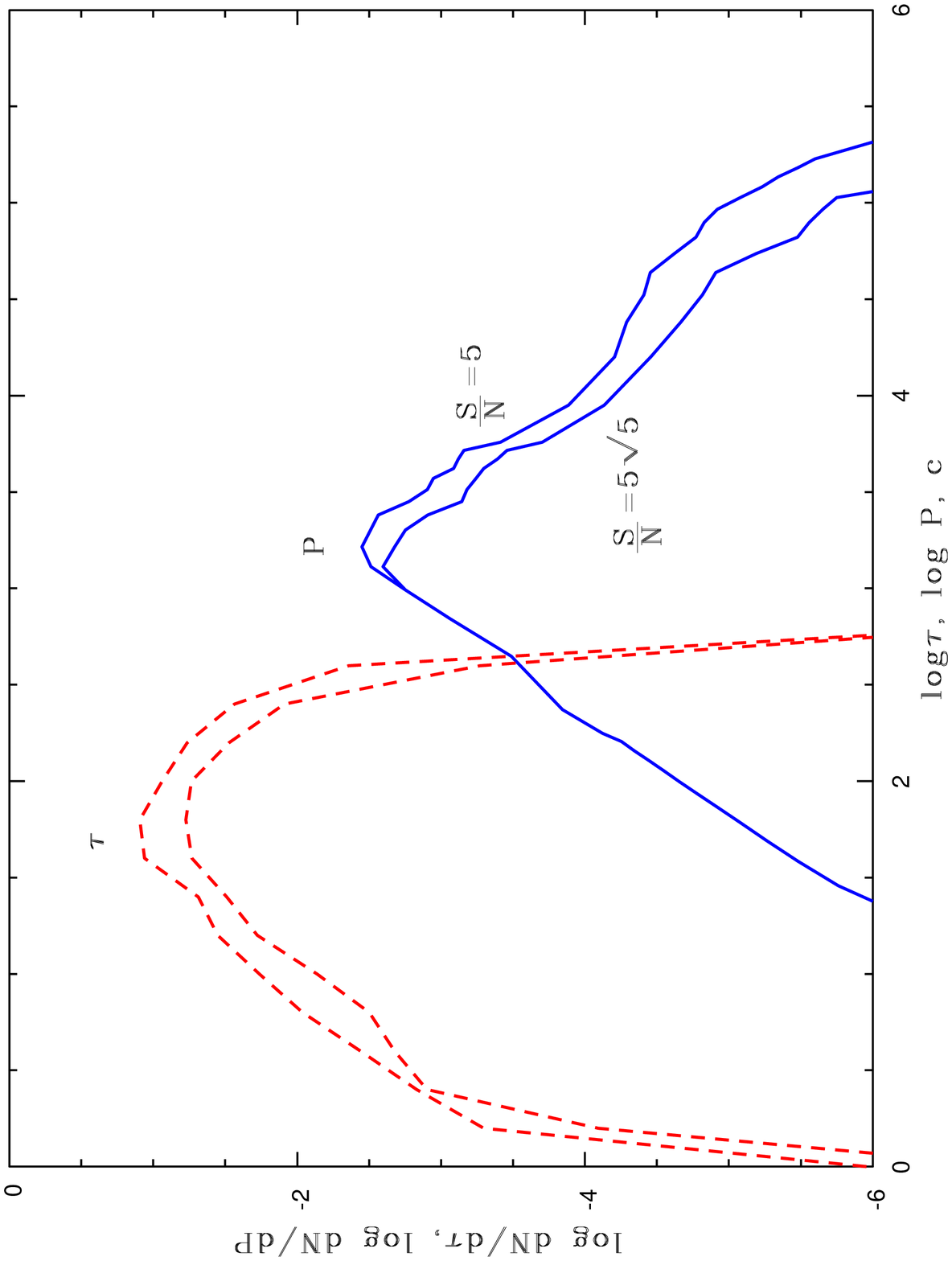}}}
\end{turn} 
\caption{ }
\label{dn_dtau}
\end{figure*}

Fig..~\protect\ref{dn_dtau}.~
The distribution of binary NSs with GW amplitudes
at periastron exceeding the LISA noise by $5\sqrt{5}$ and 5 times
(lower and upper curves, respectively), over the pulse
duration 
$\tau_{90}$ (the dashed curves) and orbital period $P$
(the solid curves). The stationary distribution function form 
Fig. \protect\ref{dn/df/de} is used. The integral over the
distribution for $S/N=5\sqrt{5}$ (lower curves) is $\sim 7$
and for $S/N=5$ (upper curves) is $\sim 15$.

\newpage

\newpage

\newpage

%


\newpage 
 
 
 
\begin{thebibliography}{50}
\bibitem{}
Arzoumanian Z., Chernoff D.F., Cordes J.M.//
ApJ, 2001, in press (astro-ph/0106159)

\bibitem{}
Bender, P.~L. et al.//~2000, LISA: Laser 
Interferometer Space Antenna. 
A Cornerstone Mission for the observation of gravitational waves -- System 
and Technology Study Report, ESA-SCI(2000)11; available at ftp://ftp.rzg.mpg.de/
pub/grav/lisa/sts-1.04.pdf

\bibitem{}
Blaauw A.// Bull. Astron. Inst.  Netherlands, 1961, 15, 265

\bibitem{}
Blinnikov S.I., Novikov I.D., Perevodchikova T.V>, Polnarev A.G.//
SvA Lett, v. 10, p.. 177


\bibitem{}
Braginsky V.B. // UFN, 2000, v. 170,p. 743.

\bibitem{}
Brown G.E., Lee C.H., Portegies Zwart S.F., Bethe H.A.//
ApJ, 2001, v.547, p.345

\bibitem{}
Buitrago J., Moreno--Garrido C., Mediavilla E.//
MNRAS, 1994,v.268, p.841

\bibitem{}
Bulik T., Belczynski K., Zbijewski W.//
MNRAS, 1999, v.309, p.629B

\bibitem{}
Doroshenko O., Loehmer O., Kramer M., et al.)//
Astron. Astrophys.2001, in press (astro-ph/0110023)

\bibitem{}
Grishchuk L.~P., Lipunov V.~M., Postnov K.~A., Prokhorov M.~E.,
Sathyaprakash B.~S.//Physics-Uspekhi,2001, v. 44. p. 1

\bibitem{}
Hellings R.W.// 2001, gr-qc/0110052

\bibitem{}
Ignatiev V.~B., Kuranov A.~G., Postnov K.~A., 
Prokhorov M.~E.// 
MNRAS, 2001, v.327, p.531

\bibitem{}
Lyne A.G., Lorimer D.R.//Nature, 1994, v.369, p.127 

\bibitem{}
Landau L.D., Lifshitz.// Field Theory. Oxford. Pergamon Press, 1969

\bibitem{}
Lipunov V.~M., Postnov K.~A., Prokhorov M.~E.//
{\it Astrophys. Space Phys. Rev.} 1996, v.9, p.1
 
\bibitem{}
Lipunov V.~M., Postnov K.~A., Prokhorov M.~E.//
MNRAS, 1997, v.288, p.245
\bibitem{}

Moreno--Garrido C., Mediavilla E., Buitrago J.//
MNRAS, 1995, v.274, p.115

\bibitem{}
Nelemans G., Yungelson L.~R., Portegies Zwart S.~F.//
Astron. Astrophys. 2001, v.375, p.890 (astro-ph/0105221)

\bibitem{}
Prokhorov M.E., Poetnov K.A.//Astron. Lett., 1997, v.23, p.492. 

\bibitem{}
Peters P.~S., Mathews J.//Phys. Rev.,~1963, v.131, p.435

\bibitem{}
Portegies Zwart S.~F., Yungelson L.~R.//
Astron. Astrophys. 1998, v.332, p.173

\bibitem{}
Portegies Zwart S.~F., Yungelson L.~R.//
MNRAS, 1999, v.309, p.26

\bibitem{}
Sosulin Yu.T.//Theoretical grounds of radiolocation and radionavigation.
Moscow, Radio i Svyaz, 1992 (in  Russian)

\bibitem{}
Tauris T.M., Sennels T.)//Astron. Astrophys., 2000, 
v.355, p.236

\bibitem{}
Taylor J.~H., Weisberg J.~M.// ApJ, 1989, v.345,
p.434

\bibitem{}
Thorne K.~S.//in {\it Three Hundred Yaers of Gravitation}
(Eds. S.~W.~Hawking, W.~Israel) (Cambridge: Cambridge Univ. Press,1987) p.330

\bibitem{}
Tikhonov V.I.// Statistical Radiophysics. Moscow, Sov. Radio, 1966
(in Russian)



\end{thebibliography}
\end{document}